\def\flash{1}
\def\aa{2}
\def\umass{3}
\def\uoa{4}
\def\so{5}
\def\lanl{6}
\def\wagce{7}
\def\sese{8}
\def\efi{9}

\newcommand{\iso}[2]{$^{#2}${#1}}
\newcommand{\msolar}{\ensuremath{M_{\odot}}}
\newcommand{\gcdsim}[1]{$63n128r{#1}z$}
\newcommand{\rhocmin}{\ensuremath{\rho_{c,min}}}
\newcommand{\sneia}{SNe~Ia}
\newcommand{\snia}{SN~Ia}
\newcommand{\gcjiv}{J08}

\documentclass{emulateapj}

\shorttitle{Detonation Mechanism of the PGCD Model}
\shortauthors{Jordan et al.}

\begin{document}

\title{The Detonation Mechanism of the Pulsationally-Assisted
Gravitationally-Confined Detonation Model of Type Ia Supernovae}

\author{
G.~C.~Jordan IV,\altaffilmark{\flash,\aa}
C.~Graziani,\altaffilmark{\flash,\aa}
R.~T.~Fisher,\altaffilmark{\umass}
D.~M.~Townsley,\altaffilmark{\uoa}
C.~Meakin,\altaffilmark{\so,\lanl}
K.~Weide,\altaffilmark{\flash,\aa}
L.~B.~Reid,\altaffilmark{\wagce,\sese}
J.~Norris,\altaffilmark{\flash}
R.~Hudson,\altaffilmark{\flash}
D.~Q.~Lamb,\altaffilmark{\flash,\aa,\efi}
}

\altaffiltext{\flash}{Flash Center For Computational Science, The
University of Chicago, Chicago, IL 60637.}

\altaffiltext{\aa}{Department of Astronomy and Astrophysics, The
University of Chicago, Chicago, IL 60637.}

\altaffiltext{\umass}{University of Massachusetts Dartmouth, Department
of Physics, 285 Old Westport Road, North Dartmouth, 02740.}

\altaffiltext{\uoa}{Department of Physics and Astronomy,
The University of Alabama, Tuscaloosa, AL 35487.}

\altaffiltext{\so}{Steward Observatory,
University of Arizona, Tucson, AZ 85721.}

\altaffiltext{\lanl}{Theoretical Division,
Los Alamos National Laboratory, Los Alamos, NM 87545.}

\altaffiltext{\wagce}{Western Australian Geothermal Centre of Excellence, CSIRO CESRE,
Kensington, WA 6151 Australia}

\altaffiltext{\sese}{School of Environmental Systems Engineering, University of Western
Australia, Crawley WA 6009 Australia}

\altaffiltext{\efi}{Enrico Fermi Institute, The University of Chicago,
Chicago, IL 60637.}

\begin{abstract}
We describe the detonation mechanism comprising the ``Pulsationally Assisted'' Gravitationally
Confined Detonation (GCD) model of Type Ia supernovae (\sneia).
This model is analogous to the previous GCD model reported in \citet{2008ApJ...681.1448J};
however, the chosen initial conditions produce a substantively different detonation mechanism,
resulting from a larger energy release during the deflagration phase.  The resulting
final kinetic energy and $^{56}$Ni yields conform better to observational values than is
the case for the ``classical'' GCD models.
In the present class of models, the ignition of a deflagration phase leads to a rising, burning plume of ash.
The ash breaks out of the surface of the white dwarf, flows laterally around the star,
and converges on the collision region at the antipodal point from where it broke out.
The amount of energy released during the deflagration phase is enough to cause the star
to rapidly expand, so that when the ash reaches the antipodal point, the
surface density is too low to initiate a detonation.
Instead, as the ash flows into the collision region (while mixing with surface fuel), the star reaches its
maximally expanded state and then contracts.
The stellar contraction acts to increase the density of the star, including the density
in the collision region.
This both raises the temperature and density of the fuel-ash mixture in the collision region
and ultimately leads to thermodynamic conditions that are necessary for the Zel'dovich gradient
mechanism to produce a detonation.
We demonstrate feasibility of this scenario with three 3-dimensional (3D), full star simulations of this
model using the FLASH code.
We characterized the simulations by the energy released during the deflagration phase, which
ranged from 38\% to 78\% of the white dwarf's binding energy.
We show that the necessary conditions for detonation are achieved in all three of the models.
\end{abstract}


\section{Introduction\label{SEC:INTRO}}

Type Ia supernovae (\sneia) are among the most energetic explosions in the known universe,
releasing $\sim 10^{51}$ ergs of kinetic energy in their ejecta, and synthesizing $\sim 0.7$ \msolar\ of
radioactive \iso{Ni}{56}.
The discovery of the Phillips relation \citep{phillips93} has enabled the use of \sneia\ as standardizable
cosmological candles which greatly enhances the accurate determination of their distance.
The discovery of the accelerated expansion of the universe using
\sneia\ \citep{1998AJ....116.1009R, 1999ApJ...517..565P} has
stimulated a tremendous amount of interest in the use of \sneia\ events
as standard cosmological candles, allowing them to serve as probes of the equation
of state of dark energy, as parameterized by the EOS parameter $w=P/\rho$.
The main challenge to the enterprise of measuring $w(z)$ using \sneia\
is reducing the systematic errors in the
accuracy with which such supernovae can be used as standard candles
\citep{detf2006}.  The accuracy must be improved from the current level
of about 15\% to about 1\% in order for large surveys
to determine the values of $w(z=0)$ and its rate of
change with $z$ to better than 10\% \citep{kim2004}. The best hope for
improvements in distance modulus accuracy is more accurate modeling of
\sneia\ explosions.

Evidence suggests that \sneia\ are the results of the thermonuclear explosion
of a carbon-oxygen (C-O) white dwarf (WD).
The leading scenario for \sneia\ explosions is the single-degenerate model
in which a progenitor WD accretes material from a non-degenerate
companion star until the mass of the WD grows nearly equal to
the Chandrasekhar limit \citep{1973ApJ...186.1007W, 1982ApJ...253..798N}.
The WD then manages to release enough
nuclear energy by fusing C and O into radioactive Ni and other lighter
$\alpha$-elements in the time span of a few seconds or less \citep{1984ApJ...286..644N},
such that it deposits approximately $10^{51}$ ergs of energy,
unbinding the star and accelerating the stellar material to speeds of thousands
of kilometers per second \citep{1982ApJ...252L..61B}.

This rapid fusion process must proceed in two phases \citep{1997ApJ...475..740N,2007ApJ...668.1132R}.
The first phase begins with the initiation of a subsonic nuclear burning
front (referred to as a deflagration or flame).
As the WD accretes material from its companion, convective carbon burn
begins in its core.
When carbon burning becomes too vigorous for convective cooling to be effective,
a thermonuclear flame (or deflagration) is born in local hot spots that form in the convective region.
The deflagration burns and rises due to buoyancy as it makes its way from the core
to the outer layers of the star.

The second phase consists of a supersonic burning front ---
a detonation --- that consumes the remainder of the WD.
The transition from the deflagration phase to the detonation phase is poorly
understood, and has been the subject of much modeling research.
A variety of models incorporates the scenario of a deflagration followed by a detonation,
such as the deflagration-to-detonation transition (DDT) \citep{1991A&A...245..114K, 2004PhRvL..92u1102G, 2005ApJ...623..337G},
the pulsating reverse detonation (PRD) \citep{2009ApJ...695.1244B,2009ApJ...695.1257B},
or the gravitationally confined detonation GCD \citep{2007ApJ...668.1118T, 2008ApJ...681.1448J, 2009ASPC..406...92J, 2009ApJ...693.1188M}.
These models differ primarily in the method by which the deflagration leads to a detonation.

An alternative scenario to the single-degenerate model is the double-degenerate model
in which two WDs forming a binary system undergo a merger process that leads
to the detonation of one of the WDs.
The double-degenerate channel \citep{1984ApJ...277..355W, 1984ApJS...54..335I}
has recently received revived attention from both observation
(e.g., \citet{2010ApJ...722.1879M}) and theory (e.g., \citet{2010ApJ...722L.157V, 2011arXiv1109.4334Z}).
However, while computational models have explored prompt
detonations in near-equal mass super-Chandrasekhar mergers \citep{2010Natur.463...61P,2011A&A...528A.117P}
and head-on collisions of binary WD systems \citep{2009MNRAS.399L.156R, 2009ApJ...705L.128R}
it remains unclear whether
more commonplace mergers between two typical C-O WDs of masses $\sim
0.6 M_{\odot}$ will result in a Type Ia explosion. Previous
one-dimensional theoretical models suggest that such mergers will
result in deflagration wave that sets off an accretion-induced
collapse to a neutron star \citep{1991ApJ...367L..19N, 1998ApJ...500..388S,
2004ApJ...615..444S, 2011arXiv1108.4036S}.

In this work we focus on the GCD scenario of the single-degenerate model of \sneia.
We use the FLASH code \citep{2000ApJS..131..273F, dubey2009} to 
extend the set of 3D whole-star GCD models in \citet{2008ApJ...681.1448J} (hereafter \gcjiv) to
include multiple ignition points as initial conditions.
These initial conditions provide more burning, and hence more
energy release during the deflagration phase.
The purpose of this work is to examine the conditions produced in the WD resulting
from a scenario where more energy is released during the deflagration phase than in
the previous GCD models and to demonstrate that the scenario manifests the necessary
conditions for a detonation.

To distinguish between the two versions of GCD models we refer to the previous body of GCD models
(\citet{2007ApJ...668.1118T},  \gcjiv, \citet{2009ApJ...693.1188M}, and references therein)
as ``classical'' GCD models and refer to models
in this paper as ``pulsationally-assisted'' GCD models or simply
``pulsational'' GCD (PGCD) models.
The origin of the name stems from the fact that the WDs in the models
undergo a pulsation where they first expand due to energy input from the deflagration phase
and then contract from the pull of gravity.
These models require the contraction of the WD to create
the thermodynamic conditions necessary to initiate a detonation,
hence the phrase ``pulsationally-assisted''.
In fact, the paradigm we put forth has characteristics of
both the classical GCD models and the PRD models.

In section \ref{SEC:GCDREV} we review the classical GCD model,
introduce the PGCD model, and discuss the ignition of the deflagration.
In section \ref{SEC:DET} we discuss the
Zel'dovich gradient mechanism as a detonation trigger
and its implications for the PGCD.
We additionally discuss the numerical detonation trigger
used in our simulations.
We give an overview of the
FLASH code and the relevant physics modules used for
our \sneia\ simulations in section \ref{SEC:METHODS}.
In section \ref{SEC:SIMS} we describe the results
of our simulations.
We compare the detonation mechanisms between the classical
GCD and the PGCD models in section \ref{SEC:COMPARE}.
Finally, in section \ref{SEC:DISCUSSION} we discuss properties
of the simulations as well as possible observational features
of the models.

\section{Discussion of GCD Models\label{SEC:GCDREV}}

\subsection{Classical GCD\label{subsec:cgcdrev}}
The classical GCD model of \sneia\ falls under the
general category of single-degenerate models.
In the GCD scenario, an off-centered deflagration ignites and begins
burning its way through the star.
The deflagration burns and rises, forming a plume of ash whose
volume is bounded by the flame.
When the plume reaches the stellar surface, the ash breaks through and
spreads around the star.
The ash then collides with itself at a location on the star
opposite that of breakout.
During this collision process, cold, low-density fuel is pushed
ahead of the ash flows.
The ash flows compress and heat the fuel in the collision region.
By squeezing the fuel in the collision region, a jet is formed.
The jet pushes the hot, smoldering fuel towards the high density
layers of the WD which leads to conditions necessary to trigger
a detonation (\gcjiv, \cite{2009ApJ...693.1188M}).
The detonation occurs between 1.5 s and 3 s after ignition, 
which is about the time it takes for the ash to flow around 
the stellar surface.

In the classical GCD scenario, the deflagration does not drive the WD 
to expand energetically due to the fact that the deflagration is so weak.
Typically, deflagration phase only release $\sim 10\%$ of the binding
energy to the WD.
The star has only mildly expanded by the time the ash flows have triggered
a detonation, and in fact the star is still expanding when the detonation is triggered.
Additionally, as reported in \gcjiv\, classical GCDs produce more than 1 \msolar\ of 
\iso{Ni}{56} which correspond to over-luminous \sneia.

\subsection{Pulsationally-Assisted GCD\label{subsec:pgcdrev}}
In this work we introduce the PGCD scenario.
The PGCD proceeds similarly to the classical GCD.
An off-centered deflagration ignites and burns its way to the surface.
The ash breaks through the surface of the star, some of which flows laterally
over the surface towards the collision region.
The PGCD differs from the 
classical GCD in the amount of energy released during the deflagration phase.
In the classical GCD scenario, a relatively small amount of energy
is released during the deflagration phase; instead, substantially more is released
during the deflagration phase in the PGCD scenario.
The WD in the PGCD scenario responds to the deflagration by expanding rapidly.
By the time the ash flows reach the collision region, the density has significantly dropped
and the flows can not compress the fuel to the high temperatures and densities
as in the classical GCD.
Thus, a detonation does not immediately occur as it did in the classical GCD.
However, as the ash flowed towards the collision region, it mixed with fuel on the
WD surface, forming a mixing layer at the interface between the ash
and surface fuel.
After the collision,
the mixture pushes its way into the
surface layers of the star.
Meanwhile, the WD has reached maximum expansion and begins to contract.
As the WD contracts, it squeezes the fuel-ash mixture
to high temperatures and densities.
The conditions obtained are above the threshold necessary to trigger a detonation.
Thus, it is the stellar contraction, and not just the kinetic energy in the 
ash flows themselves (as in the classical GCD), that gives rise to conditions
that make it possible for the mixture to detonate.

\subsection{Ignition of the Deflagration\label{subsec:ics}}
The simulations of the classical GCD model in \gcjiv\ were initiated with a single
ignition point that was offset from the center of the WD.
These initial conditions lead to a weak deflagration
which gives rise to the events described in section \ref{subsec:cgcdrev}.
We extended our set of initial classical GCD models by choosing initial conditions
that would lead to more burning --- and thus releases more energy --- than we had previously obtained.
To accomplish this we chose to initialize our simulations with a single cluster of multiple
ignition points that was offset from the center of the WD.
These initial conditions serve to represent two physical situations that may
occur in the WD.
First, it may be possible that ignition occurs at multiple points.
Recent work by \citet{2011ApJ...740....8Z} and \citet{2012ApJ...745...73N}
of the last moments of the smoldering phase before ignition conclude
that a single ignition point is the most probable scenario; however,
ignition is an inherently stochastic process and many realizations are required
to make definitive statements about all of the possible outcomes.
Currently only a handful of simulations exist.
Furthermore, the process of forming hot spots in the simulations that
eventually lead to an ignition may be resolution dependent as stated
in \citet{2012ApJ...745...73N}, and leaves open the door for a multiple
ignition point scenario.
Finally, all but one of the  simulations in \citet{2011ApJ...740....8Z} and \citet{2012ApJ...745...73N}
were stopped once a hot spot formed that lead to ignition.
Therefore, it may have been possible for more ignition points to form
after the initial runaway, even though in the specific simulation  that they continued
past the formation of initial ignition point no secondary ignition points
were observed.
Second, though we ignore the convective turbulence, most likely
convective plumes exist with a velocity $\sim 100$ km/s \citep{2012ApJ...745...73N}.
This is comparable to the flame speed and can potentially
grow the flame surface through the Kelvin-Helmholtz instability.
We essentially mock-up this effect by initializing our simulations with multiple ignition points
in close proximity to one another.
These ignition points in turn quickly merge and form a single asymmetric burning
region enclosed by a distorted flame surface.

\section{Detonation Mechanism\label{SEC:DET}}
\subsection{Theory \label{subsec:theory}}
\subsubsection{Discussion of the Zel'dovich Gradient Mechanism}
In the PGCD model, 
a plume of ash, with a layer mixed with fuel, plunges into the star
during its contraction phase
and initiates a detonation.
Initially, as the plume advances around the star, a mixing layer of cold fuel and hot ash forms
at the star-ash interface.
Compression due to the stellar contraction and to the flow pushing to higher
density layers of the star acts to heat the mixture.
The composition in the mixing layer transitions from pure, hot ash in the plume to pure, cold fuel
in the star, resulting in a compositional and temperature gradient in the mixing layer.
A spontaneous detonation is triggered
by the Zel'dovich gradient mechanism \citep{1970JAMTP..11..264Z} when the
mixture reaches the critical conditions.

The Zel'dovich mechanism requires an induction time gradient resulting
from a gradient in temperature or gradients in temperature and composition.
Rapid combustion in the high temperature region results in a shock wave as
the burning front progresses to regions of low temperature.
A detonation is formed if the gradients are shallow enough and the
combustion becomes associated with the shock;
otherwise, the shock wave runs ahead of the burning front and the plasma does not
detonate.

For the purposes of this work, we wish to characterize the conditions
under which a detonation would occur in the mixing layer that is formed
when the ash plume flowed from its breakout point until it reaches the
collision region.
In the envisaged scenario, the mixing layer consists of a compositional
gradient between pure ash in the plume and pure fuel on the surface of the WD,
as well as a temperature gradient produced by combining the varying amounts
of hot fuel and cold ash.
In the following we discuss the critical conditions that are necessary to trigger a detonation
in such a scenario.
In particular, we concentrate on the possibility that the gradient mechanism
would trigger a detonation on length scales that are under resolved or unresolved in our simulations,
which corresponds to length scales $\lesssim 10$ km.

Much work has been done to determine the properties of the gradients required
to successfully initiate a detonation by the Zel'dovich gradient mechanism \citep{1994ApJ...427..330A, 1997ApJ...475..740N,
1997ApJ...478..678K,2007ApJ...660.1344R, 2009ApJ...696..515S}.
Both \citet{2007ApJ...660.1344R} and \citet{2009ApJ...696..515S} studied systems
initialized with a temperature gradient that included a constant density and composition (though
the PGCD would also have a compositional gradient).
\citet{2007ApJ...660.1344R} examined the results of a linear
temperature gradient.
\citet{2009ApJ...696..515S} performed a more comprehensive
study with several functional forms for the temperature gradient and a wider range
of conditions.
The goal of the work was to determine the
minimum size of the region containing the induction
gradient for a self sustaining shock-reaction complex to form.
This minimum size (or critical length scale, $L_{c}$) was determined to be a function
of the input parameters of the system such as the ambient temperature,
ambient density, peak temperature, steepness and functional form of the temperature profile,
and composition of the fuel.
In general, they find that for $L_{c} \lesssim 10$ km, a composition of 50\%-50\% \iso{C}{12}\ and \iso{O}{16}\ with
densities $\gtrsim 1 \times 10^{7}$ g/cm$^3$ and peak temperatures $\gtrsim 2 \times 10^{9}$ K would successfully trigger
a detonation.

\citet{2009ApJ...696..515S} performed a resolution study on one of their models
and examined how $L_{c}$ changed when they increased their resolution.
In this study, the $L_{c}$ grew from $\sim 400$ m in the lowest resolution case to $\sim 1,000$ m in the highest resolution
case (see their table 3).
The grid resolution with which they ran their simulations may increase $L_{c}$ for
a given parameter set, or said another way, the grid resolution will increase the critical
conditions required to trigger a detonation for a particular $L_{c}$.
Because they only performed the resolution study on one set of parameters it is difficult to asses
how much the critical conditions would increase with adequate resolution for a given $L_{c}$.

\citet{2009ApJ...696..515S} also examined the conditions for the formation of a detonation
for systems comprised of less than 50\% \iso{C}{12}\ by mass.
In their most extreme case of carbon-poor fuel (30/70 C/O ratio) the critical radius is roughly
a factor of 10 larger than fuel composed of equal amounts of carbon and oxygen.
In general, the reduction of \iso{C}{12}\ increased the critical length scale (their tables 8 - 11).
They also tested the effect of adding \iso{He}{4}\ (14\% by mass) to the composition.
The addition of helium dramatically decreased the critical length scale (their table 12).
As with the resolution study, they only examined one set of parameters for their composition study,
and thus did not determine how changing the composition with a fixed $L_{c}$ would change
the values of peak temperature, density, etc., required for a successful detonation.
We note that by lowering the \iso{C}{12}\ mass, the critical conditions would most likely
would increase, and for extremely carbon-poor environments a detonation may not be triggered at all.
For the remainder of this work we assume that extremely carbon-poor environments are not encountered and
that small to moderate deviations from a 50/50 C/O ratio would not substantially increase the critical
temperature and density --- at least to within the context of our approximation.

\citet{1997ApJ...478..678K} examined the conditions under which the
Zel'dovich gradient mechanism would produce a detonation given
a gradient in both temperature and composition.
They study two cases:
In the first case turbulence tears apart an active flame and
mixes cold fuel with hot ash to achieve
the necessary conditions to initiate a detonation.
In the second case, stellar expansion extinguishes the active flame
allowing the hot ash and cold fuel to mix.
As the star contracts, it squeezes the mixture - creating
the necessary conditions for the mixture to detonate.

\citet{1997ApJ...478..678K} argue that mixing is important in order for fuel to attain
a high enough temperature to ignite at relatively low densities.
They point out that cold fuel that obeys a WD equation of state 
would have to be compressed to $\approx 1 \times 10^{10}$ g/cm$^3$
in order to obtain a high enough temperature to ignite; however,
mixing hot ash at a level as low as 10\% raises
the entropy enough that the mixture would ignite at densities $\approx 10^7$ g/cm$^3$.
Their figure 2 shows the critical density for ignition as a function
of fuel mass fraction in the mixture.

In \citet{1997ApJ...478..678K}, a temperature gradient is set up in the mixing layer between cold fuel and hot ash.
The temperature of the mixture is low in regions dominated by fuel and is high in regions
dominated by ash.
It is thus a matter of compressing the mixture to the critical density (and thus the critical
peak temperature through compressional heating) for a detonation to initiate in the mixture. 
Their Figure 6 shows $L_{c}$ as a function of critical density for a gas initialized with a linear
compositional gradient and a temperature gradient determined by the amount of ash in the mixture at
a particular location.
They conclude that in the WD environment,
a detonation is possible for densities between $5 \times 10^{6}$ g/cm$^3$ and
$2-5 \times 10^{7}$ g/cm$^3$.
For a mixing layer with a critical length scale on the order of 10 km,
the critical density of the fuel is $\approx 1 \times 10^{7}$ g/cm$^3$.
Interestingly, this is commensurate with the results of \citet{2009ApJ...696..515S}
with a  50/50 C-O composition, even though the two approaches were distinct.

We therefore conclude that for $L_{c} \approx 10$ km,
the critical density, $\rho \gtrsim 1 \times 10^{7}$ g/cm$^3$ and
with a corresponding critical peak temperature, T $\gtrsim 2 \times 10^{9}$ K
are the necessary criteria to successfully trigger a detonation on that length scale.
We remind the reader that we have assumed that the environment is not carbon-poor
and that a decrease in the fuel mass will not dramatically increase the critical
conditions.
We have further assumed that though the method used to determine the critical values
was unresolved, the results of a fully resolved study would not dramatically increase the critical
values from the unresolved case.
Though both the composition effects and uncertainties with the resolution of the
method will affect the values critical for detonation, the values of temperature
and density that we have selected are acceptable for our level of approximation.

\subsubsection{Discussion of Mixing at the Fuel-Ash Interface}
The PGCD detonates in the same manner as the second
case described in \citet{1997ApJ...478..678K}, which depends on the formation
of a fuel-ash mixture.
In the PGCD, ash ejected from the star flows over the surface to the collision
region.
As the ash flows, it turbulently mixes with fuel on the surface forming
a mixing layer of fuel and ash.
In the following we demonstrate the feasibility of our mixing assumption by examining the
growth of Kelvin-Helmholtz modes at the fuel-ash interface and showing
that these modes can grow quickly enough and on the relevant length scales
to provide the appropriate mixing.
However, mixing is a complicated process and our assumptions below of constant
physical conditions are meant to provide an order-of-magnitude estimates.
We are particularly interested on the growth of the mixing layer on unresolved scales.
As above, our length scale of interest is $L_{c} \approx 10$ km.

Using typical values from the conditions present at the surface-ash interface,
we can approximate the largest unstable Kelvin-Helmholtz length
scale, $\lambda_{max}$, such that all $\lambda \le \lambda_{max}$ would
grow due to the instability.
Using linear stability analysis, the largest unstable length scale is expressed in terms of the physical
conditions at the interface between the flowing ash and the surface of the
WD as 
\begin{equation}
\lambda_{max} = { 2 \pi \cdot \alpha_{fuel}  \alpha_{ash} U^{2} \over g (\alpha_{fuel} - \alpha_{ash})},
\label{eq:kh}
\end{equation}
where $\alpha_{fuel} = {\rho_{fuel} \over \rho_{fuel} + \rho_{ash}}$,
$\alpha_{ash} = {\rho_{ash} \over \rho_{fuel} + \rho_{ash}}$,
$\rho_{fuel}$ is the fuel density (i.e. the surface density of the WD),
$\rho_{ash}$ is the ash density,
$g \approx 1 \times 10^{9}$ cm/s$^{2}$ is the acceleration due to gravity,
and $U$ is the velocity of the ash \citep{1961hhs..book.....C}.
The ash flows at a speed $U \approx 5\times 10 ^ {8}$ cm/s and the interface is located
at $r_{interface} \approx 5\times 10 ^ {8}$ cm from the center of the star.
Values for the densities are $\rho_{fuel} \approx 1\times 10^{6}$ g/cm$^3$ and
$\rho_{ash} \approx 1\times 10^{5}$ g/cm$^3$.
Inserting these values into equation.~(\ref{eq:kh}) gives $\lambda_{max} \approx 1.0 \times 10^8$ cm - comparable
to the radius of the star and hence $>>$ $L_{c}$ .
Therefore a region the size of $L_{c}$ would develop Kelvin-Helmholtz instabilities which would drive
turbulent mixing.

The ash must flow approximately half way around the star to reach the collision region.
Using the values of $U$ and $r_{interface}$ we can approximate the time it would take for
the ash to reach the collision region, $\tau_{ash}$, as
\begin{equation}
\tau_{ash} = {\pi \cdot r_{interface} \over U}
\label{eq:tash}
\end{equation}
Therefore $\tau_{ash} \approx 1$s.
Note that this is consistent with the results of our simulations discussed in section \ref{subsec:simnumbers}
and shown in figure \ref{fig:thetaAshVsTime}.

The timescale for the growth of the unstable Kelvin-Helmholtz modes with length 
scale $\lambda$ \citep{1961hhs..book.....C} in the linear regime is
\begin{equation}
\tau_{KH}(\lambda) = [g(2 \pi / \lambda)(\alpha_{fuel} - \alpha_{ash}) - (2 \pi /  \lambda)^{2}\alpha_{fuel}  \alpha_{ash}  U^{2}]^{-1 / 2}.
\label{eq:tkh}
\end{equation}
The growth time scale for a perturbation of length scale $\lambda = L_{c}$ km is
$\tau_{KH}(10km) \approx 1 \times 10^{-3}$ s.
Since $\tau_{ash} >> \tau_{KH}(10km)$, perturbations of length scale $L_{c}$ will
grow for several (linear analysis) e-folding times before the ash reaches the collision region.

We again note that the conditions at the ash-star interface are not constant, especially since the
star is expanding as the ash flows over the surface.
However, given the short growth timescale ($\tau_{KH} \approx 1 \times 10^{-3}$ s) for $L_{c} = 10$ km,
if the above conditions exist for only a small fraction of the time that the ash is flowing over the surface
then the requisite modes would grow and mixing would develop by the time the ash reaches the collision region.

We conclude that it is possible for the Zel'dovich mechanism to trigger a detonation in the mixed fuel-ash layer
on length scales not adequately resolved in our simulation if the critical conditions are reached.
The ash flows turbulently mix with fuel on the surface driven by Kelvin-Helmholtz instabilities.
The mixing process produces the necessary thermal and compositional gradients and the increase
in density of the mixture as it pushes into the star both accelerates the combustion and shrinks
the critical length scale, $L_{c}$, required for the gradients.

\subsection{Numerical Treatment of Detonation Trigger\label{subsec:numreal}}
To treat the detonation physics we have incorporated the numerical scheme
used by \citet{2009ApJ...693.1188M} and \citet{2009ApJ...696..515S} which
we briefly describe in section \ref{SEC:METHODS} and which was not present in \gcjiv.
To correctly capture the Zel'dovich mechanism in the simulation
we would need to resolve the required gradients as well as the carbon burning length scales
within those gradients.
This is prohibitively expensive for 3D simulations and we do not accomplish it here.
Thus the initiation of a detonation in our simulations is a process that is severely under resolved.
The finest resolution in our simulation does not allow us to follow the simulation on length scales of $\lesssim 10$ km.
In section \ref{subsec:theory} we estimate that if a gradient is characterized by a peak temperature, T $\gtrsim 2 \times 10^{9}$ K and
density,  $\rho \gtrsim 1 \times 10^{7}$ g/cm$^3$ on length scales of $10$ km then the necessary conditions exist to produce a gradient-triggered
detonation.
We further show that mixing could occur on length scales on the order of $10$ km as the ash flows
around the star.
Thus we adopt the criteria that if a computational cell's temperature and density exceed those specified
above, then the necessary conditions exist for a gradient-triggered detonation and we allow the cell
to detonate.

Numerically, we trigger the detonation by allowing the \iso{C}{12} - \iso{C}{12}\ reaction
to run away.
When a computational cell
reaches a high enough temperature
and density such that the carbon burning reaction rate becomes extremely high,
all of the fuel in the cell burns in a single computational time step.
If there is enough fuel in the cell, this releases a tremendous amount of energy, which dramatically
increases the temperature and pressure in the cell, and creates a shock
with the neighboring cells.
The shock propagates outwards from the cell and burns material in its wake, forming
a detonation.
This is akin to the direct initiation of a detonation by a blast wave.
Since the initiation of the detonation is purely numerical in nature, we stress that
it simply indicates that a computational cell has reached the necessary thermodynamic
conditions for rapid carbon burning.
However, the conditions for rapid carbon burning also coincide with the necessary (but not sufficient) conditions
that we expect would enable the Zel'dovich mechanism to produce a detonation as outlined in section \ref{subsec:theory}.
We refer the reader to \citet{2009ApJ...693.1188M} and \citet{2009ApJ...696..515S,2009ApJ...700..642S} for a
discussion of the initiation of detonations in the context of numerical simulations.

In our description of the detonation mechanism, we appeal to the fact
that the ash ejected from the WD will mix with fuel on the surface
of the star as it flows; however, we do not treat mixing on unresolved scales.
Mixing on small scales occurs by numerical diffusion.
Though we do not capture the details of that mixing process, 
we fully expect the material to be well mixed as it flows around
the star as postulated in section \ref{subsec:theory}.

Finally, due to the above mentioned physical processes that are not included
in our simulations, the physical quantities that make up the trends discuss in section \ref{SEC:SIMS} and
listed in table \ref{tab:sims} that are the results of the detonation
should be taken with a grain of salt.
These properties (such as the radius at which the detonation occurs for example)
are the result of the time at which the numerical detonation is triggered and
the state of the contracting WD at that time.
Though the conditions for which the numerical detonation is triggered roughly coincides
with those that are sufficient to trigger a gradient-induced detonation, by approximating the relevant physics
the exact time and location of the detonation are uncertain.
Altering the detonation time and location could change the physical properties resulting from
the detonation.
Nevertheless because of the correspondence between the conditions for the numerical detonation
and that of the physical detonation, we expect the general trends presented in section \ref{SEC:SIMS}
to hold while we acknowledge that there may be large uncertainties.

\section{Numerical Methods\label{SEC:METHODS}}

We use the Adaptive Mesh Refinement (AMR) FLASH application framework \citep{2000ApJS..131..273F,dubey2009}
to perform the simulations presented in this paper.
FLASH is a modular, component-based application code framework created
to simulate compressible, reactive astrophysical flows.
The framework supports multiple methods for managing the discretized
simulation mesh, including the PARAMESH (Parallel Adaptive Mesh Refinement)
library \citep{2000CoPhC.126..330M}, which implements a block-structured
adaptive grid.

FLASH includes a directionally split piecewise-parabolic method (PPM)
solver \citep{1984JCoPh..54..174C} descended from the PROMETHEUS code \citep{1989ApJ...341L..63A}.
%
It has been successfully applied to a wide variety of large-scale terrestrial
and astrophysical flow problems, ranging from simulations of
homogeneous, isotropic turbulence \citep{2008PhRvL.100w4503B, 2008PhRvL.100y4504A},
Raleigh-Taylor instability \citep{2002ApJS..143..201C},
shock-cylinder interaction \citep{2005Ap&SS.298..341W},
and laser-driven high energy density laboratory experiments \citep{2001PhRvE..63e5401K},
to buoyancy-driven turbulent combustion \citep{2008JPhCS.125a2009T},
wind-driven instabilities in neutron star atmospheres \citep{2004ApJ...602..931A},
contact binary stellar evolution \citep{2008ApJ...672L..41R},
Type Ia supernovae \citep{2008ApJ...681.1448J,2009ApJ...693.1188M,2009ASPC..406...92J},
galaxy collisions \citep{2008AAS...212.2107Z},
and cosmological simulations of large-scale structure formation \citep{2005ApJS..160...28H}.

The energetics scheme employed to treat flames and detonation waves in our simulations uses three progress
variables to track carbon burning, QSE relaxation, and NSE relaxation.
Details concerning the nuclear physics and the numerical implementation
are presented in \citet{2007ApJ...656..313C, 2007ApJ...668.1118T} and \citet{2009ADNDT..95...96S}.
Subsonic burning fronts (deflagrations) are advanced using an advection-diffusion-reaction (ADR) equation.
Details concerning the implementation, calibration and noise properties of the 
flame treatment can be found in \citet{2007ApJ...668.1118T} and references therein.
Detonations are handled naturally by the reactive hydrodynamics solver in FLASH
without the need for a front tracker.
This approach is possible because unresolved Chapman-Jouguet (CJ) detonations retain
the correct jump conditions and propagation speeds.
Cellular structure smaller than the grid scale will be suppressed in our simulations but is free
to form on resolved scales.
The impact of cellular structure on the global evolution of the model is still uncertain;
however, since cellular structure alters the detonation wave speed by only a few percent
for the conditions being modeled \citep{2000ApJ...543..938T}, the effect is likely to be small.
Additional details related to the treatment of detonation waves are discussed in \citet{2009ApJ...693.1188M}.

Self gravity is calculated by expanding the mass density field in multipole moments,
which are used to approximate the scalar gravitational potential.
The gravitational acceleration is calculated by approximating the
derivative of the scalar potential at each location in the domain.
The Helmholtz equation of state of \citet{2000ApJS..126..501T} is 
used to describe the thermodynamic properties of the stellar plasma, including contributions 
from blackbody radiation, ions, and electrons of an arbitrary degree of degeneracy.

\section{Simulations and Results\label{SEC:SIMS}}
\subsection{Simulation Setup\label{subSEC:SIMSet}}
We extended the study of the GCD model described in \gcjiv\
with two primary differences.
First, we included detonation physics in the simulations (as we described in
sections \ref{subsec:numreal} and \ref{SEC:METHODS})
and followed the models from
ignition, through the detonation phase, and to the free expansion phase.
We terminated the simulations when the temperature
decreases to the point that nuclear reactions ceased.
Second, we initiated these simulations with multiple ignition points (as described in section \ref{subsec:ics})
instead of a single ignition point with the method used in \gcjiv.
For completeness we give the basic points here but refer the reader to
\gcjiv\ and references therein for more details.

Our simulations used FLASH's AMR capabilities with a finest resolution of 8km.
Each simulation contained a 1.365 \msolar\ WD in hydrostatic equilibrium
composed of equal parts carbon and oxygen.
The WD had a central density of $2.2 \times 10^9$ g/cm$^{3}$, an ambient
temperature of $3 \times 10^{7}$ K, a radius of approximately $2000$ km,
and a binding energy of $4.9 \times 10^{50}$ ergs.
Although the core of the star is most likely convective
with a turbulent
convective RMS velocity $v_{\rm RMS} \sim 16$ km/s \citep{2012ApJ...745...73N},
the turbulent
convective RMS velocity is much less than the laminar flame speed, and
so we have ignored the background convective turbulence,
and initialized the star with zero velocity.

We initiated this series of simulations with multiple ignition ``points'',
which are $16$ km spheres comprised of hot ash placed in the star at rest.
We placed the ignition points in a spherical region whose center
coincided with the z-axis.
We parameterized the spherical region by its radius and z-offset (the distance from the origin of the
center of the spherical region along the z-axis).
The radius of the spherical region was $128$ km and
the z-offsets were $68$ km, $88$ km, and $168$ km - one z-offset for each of our three simulations.
We randomly populated a $128$ km spherical region with $63$ ignition points.
We placed ignition points so that they did not overlap with one another, and so that they were contained
entirely within the spherical region.
We used the same random distribution of ignition points for all three simulations;
they only differed by their relative location along the z-axis.
Table \ref{tab:sims} contains labels for the simulations referred to in this work along
with parameters for the initial conditions.

\subsection{Evolution of Simulations\label{subsec:simdescript}}

In this section we describe the evolution of the PGCD model as demonstrated
by the \gcdsim{168} simulation.
Figures \ref{fig:gcdpBrads1} and \ref{fig:gcdpBrads2} show snapshots of the
\gcdsim{168} simulation starting with the initial conditions and ending with the
passage of the detonation wave.
The green contour in the figures approximates the location of the WD's surface.
The orange regions are high temperature regions.
These regions are primarily hot ash.
The temperature ranges from $1 \times 10^9$ K (dark orange) to $5 \times 10^9$ K (bright orange).

Each simulation begins with one of the above multiple ignition point configurations.
Panel ($a$) of Figure \ref{fig:gcdpBrads1} shows the initial conditions of
the \gcdsim{168} simulation.
In the first few tenths of a second, each ignition point quickly burns
radially outward from its center and merges with the other ignition points to form a
large asymmetric bubble of ash.
The ash in the bubble is less dense than the surrounding stellar material and is therefore buoyant.
After 0.3-0.4 seconds, the bubble, which continues to grow in size, begins to quickly rise towards
the WD's surface.  
During the rise, Rayleigh-Taylor instabilities grow on the flame surface,
which further increases the surface's complexity and
enhances the burning rate \citep{1995ApJ...449..695K,2008JPhCS.125a2009T}.

Panel ($b$) of Figure \ref{fig:gcdpBrads1} shows the simulation at 0.6s.
This panel shows the ignition points after they have risen and merged to
form a complex volume whose surface has been enhanced by fluid instabilities.

After approximately 1.5 s, the rising ash bubble breaks through the stellar surface 
and begins spreading laterally across the star in all directions, converging on a region antipodal
to the bubble breakout region.
Since the density of the surface layers of the star are too low to maintain
thermonuclear combustion, the flame quenches and the deflagration subsides.
As the ash flows over the surface of the star, cold fuel (C and O) mixes with the ash
at the interface between the ash and the stellar surface, creating a fuel-ash mixture.

Panels ($c$), ($d$), and ($e$) of Figure \ref{fig:gcdpBrads1} are at 1.1 s, 1.4 s, and 1.9 s, respectively.
These panels show the ash after it has broken out of the surface and has started to spread.
In panel ($c$) the ash has spread approximately half way around the star.
By 1.4 s (panel $d$), the ash has made it three-fourths the way around the star,
and at 1.9 s (panel $e$) the ash has almost completely engulfed the WD.

A significant amount of nuclear energy, $1.9 \times 10^{50}$ ergs
(comparable to the binding energy of the WD),
is released during the deflagration.
This provides a kick to the star which causes it to expand.
During the first few seconds, while the ash flows over the surface, the star expands to several times its
original size.
The expanding star slows the ash before it reaches the collision
region.
This shown in panels ($c$) - ($f$) of Figure \ref{fig:gcdpBrads1} and
panel ($a$) of Figure \ref{fig:gcdpBrads2}.

Panel ($f$) of Figure \ref{fig:gcdpBrads1} is at 2.2 s and
panel ($a$) of Figure \ref{fig:gcdpBrads2} is at 2.5 s.
The ash seems to disappear from the panels because the
star has expanded and the ash has cooled to the point that
it has fallen out of the color range.
Panel ($b$) of Figure \ref{fig:gcdpBrads2} is at 3.1 s.
The WD is maximally expanded in this image.

Several seconds after the ash is ejected from the surface of the star, it converges at the opposite pole
from where is broke out.
The converging flow compresses the material in the collision region until the
pressure is sufficient to stop the mixture from flowing laterally
and to drive a plume of fuel and ash towards the interior of the star.
Additionally, after the star has reached its maximally expanded state,
it begins to contract.
The stellar contraction increases the global density structure of the star,
and thus the local density of at the collision region, eventually resulting in
explosive C and O burning.
Once the conditions for detonation are met (as discussed in section \ref{subsec:theory})
and the detonation is triggered (as discussed in section \ref{subsec:numreal})
the combustion immediately sweeps across the star
in a few tenths of a second.

Panels ($c$) and ($d$) of Figure \ref{fig:gcdpBrads2}, at 3.7 s and 4.2 s respectively, show the star as it contracts.
A hot region can be seen forming at the ``bottom'' of the star.
This hot region is the result of the compressional heating of the mixture in the collision region
from the work done by the contracting star as well as the material flows themselves.
Panel ($e$) of Figure \ref{fig:gcdpBrads2} is at 4.3 s and shows the simulation just after the initiation
of the detonation wave.
Panel ($f$) of Figure \ref{fig:gcdpBrads2} shows the simulation at 4.4 s.
The detonation wave has consumed about half of the star in just over 0.1 s.

The detonation wave transforms the WD into a super-heated remnant composed of
material that has reached nuclear statistical equilibrium (NSE)
in the core (which is dominated by iron-group elements, most of which is \iso{Ni}{56})
surrounded by a layer of intermediate mass elements and topped of with
the ash of partially burned C and O.
This whole structure is shrouded by the products of the deflagration which consisted of
iron-group elements, intermediate mass elements, carbon burning products, and unburned C and O.
The super-heated structure quickly expands and cools,
and is transitioning into a homologous structure when the simulations
are stopped.

\subsection{Results\label{subsec:simnumbers}}
In each of our simulations, we found the necessary conditions to trigger a detonation
as discussed in section \ref{SEC:DET}.
The main properties of the three simulations are summarized in Table \ref{tab:sims}.
The simulations differ in the location of the center of their sphere of ignition
points, and are labeled \gcdsim{68} (ignition sphere centered 68 km from the
center of the star), \gcdsim{88} (88 km from the center of the star), and
\gcdsim{168} (168 km from the center of the star).

The deflagration phase of the \snia\ provides a kick to the
WD which causes the WD to rapidly expand and then contract.
It is the expansion and subsequent detonation on contraction that
characterizes the pulsation GCD model.
Figure \ref{fig:enucVsTime} shows the amount of nuclear energy release as a
fraction of the binding energy of the star as a function of time for the three simulations.
The simulations trend together for the first 0.8s and then diverged.
The deflagration continued to release nuclear energy until approximately 1.5s
in each simulation, at which time the flame stopped burning.
In general, the simulations whose ignition points were placed closer to the core
of the WD burned more during the deflagration and thus imparted more
energy to the star.
The amount of energy released ranged from $18.9\times 10^{49}$ ergs to $38.6\times 10^{49}$ ergs,
or between 38\% and 78\% of the binding energy of the star.
By comparison, the single bubble initial conditions in \gcjiv\ released between
$3.0\times 10^{49}$ ergs gand $10.5\times 10^{49}$ ergs, or between 6\% and 21\% of the binding energy of
the WD.

The energy released during the deflagration phase was sufficient to cause the star to undergo an
energetic pulsation.
Figure \ref{fig:rhocVsTime} shows the evolution of the maximum density found in the computational domain, 
which is equivalent to the central density of the WD (prior to detonation) for
the three simulations and gives a measure of the strength and the period of the pulsation.
In each simulation, the WD expanded to a maximum amplitude and then contracted
before detonating.
The simulation that expanded the least was \gcdsim{168} which reached a minimum 
central density, \rhocmin,  of $3.51\times 10^{7}$ g/cm$^3$.
The star expanded for 3.1 s before it began
to contract.
By contrast, \gcdsim{68} achieved $\rhocmin=0.43\times 10^{7}$~g/cm$^3$,
and the WD did not begin to contract until 4.71 s.
Thus the more energy released during the deflagration phase,
the more the WD expands and the longer its pulsational period.
The vertical lines in Figure \ref{fig:rhocVsTime} mark the time at which the
star detonates.

As the WD expanded, ash was ejected from the surface
and flowed laterally over the star, mixing with surface fuel as it flowed.
Figure \ref{fig:thetaAshVsTime} plots the polar angle of the leading edge of
the flow as a function of time.
Initially a small region of ash crossed the origin and was responsible
for the large values of $\theta$ seen in the figure during the first
second.
After one second, the ash reached the surface and started to spread around the
star.
The ash, which mixes with fuel on the surface as it flows,  quickly reached a polar angle of approximately $150^{\circ}$
and then stalled in each of the simulations.
This was in part because the mixture pushed some material in front of it
which compressed material in the collision region and increased
the pressure there.
Also, the expansion of the star robbed kinetic energy from the flow which also contributed
to the mixture stalling.
Once the star contracted, the mixture slowly pushed its way further around
the star, as well as into the high density regions towards the core.
As before, the lines on the graph highlight the time at which the star
detonated.
The mixture only stalled for a short period of time in the \gcdsim{168} simulation
whose WD had the shortest pulsational period.
The opposite is true for the \gcdsim{68} simulation in which the mixture stalled the longest and
the WD had the  longest pulsational period.
Even though the mixture encroached on the collision region within a 
few seconds from when the flow started, it was the contraction of the
star that assisted the mixture in making the final move to higher densities. 

After the star began to contract, the fuel-ash mixture made its way into
the high-density layers of the WD.
Figure \ref{fig:maxTempVsTime} shows the evolution of the thermodynamic conditions
in the fuel-ash mixture as it pushed into the star.
The top plot in figure \ref{fig:maxTempVsTime} shows the temperature of the computational
cell with the maximum temperature in the mixture.
The middle plot shows the density and
the bottom plot shows the radius of the computational cell with the maximum temperature.
Both the temperature and density follow the trend of the central density, as the WD expanded and contracted.
As the star expanded, the temperature and density of the mixture decreased.
As the star contracted, the temperature and density of the mixture increased.
The location of the hot spot followed the same pattern; hence,
as the star contracted, the radius of the hot spot moves closer to the core.
This was due both to the stellar contraction and the continued flow of the mixture.
Once the temperature of the hot spot exceeded $\sim 2.0 \times 10^9$ K and the 
density exceeded $\sim 1 \times 10^7$ g/cm$^3$ rapid combustion ensued,
which transitioned into a detonation.
We stress that the detonation is the result of the numerical scheme discussed
in section \ref{subsec:numreal} and is not the result of the Zel'dovich gradient
mechanism.
It merely indicates that the conditions in the computational cell surpass
those that are necessary to produce a gradient-induced detonation.
In the \gcdsim{168} simulation, the deflagration phase released the least
amount of energy of all the simulations and thus the
WD expanded the least, it had a shortest
pulsational period, it began contracting the earliest.
As a result, the detonation occurred the soonest.
The opposite is true for the \gcdsim{68} simulation whose deflagration phase released
the most amount of energy and whose detonation was triggered last.

An important set of observables of an \snia\ are the nucleosynthetic yields
produced in the event.
We have approximated the yields from the three models using the reaction
progress variables from FLASH's burning module \citep{2007ApJ...656..313C}.
Table \ref{tab:det} lists the post-explosion nucleosynthetic yields from our simulations in terms
of the quasi-static equilibrium groups that are represented by the progress
variables.
Note that the material that burned to NSE is
predominately iron-group elements, most of which is \iso{Ni}{56}.
Thus, the amount of NSE material can be considered an upper limit on the amount of \iso{Ni}{56}
produced during the explosion.
The upper limit on the \iso{Ni}{56} produced during these simulations ranges from $\sim 1.0$ \msolar\ to $\sim 0.7$ \msolar.
Though yields of $\sim 1.0$ \msolar\ of \iso{Ni}{56} are associated with luminous \sneia,
yields of $\sim 0.7$ \msolar\ of \iso{Ni}{56} are associated with \sneia\ of normal luminosity.
The combined mass of the intermediate mass elements and NSE material is $~1.1$ \msolar\ and is consistent
with observed \sneia\ \citep{2007Sci...315..825M}.

The deflagration phase of the PGCD models burns approximately 0.1 - 0.25 \msolar\ of material.
This material is ejected into the outer regions of the system, surrounding the WD when it
detonates.
Table \ref{tab:def} lists the nucleosynthetic yields produced during the deflagration phase of the
three simulations.
The primary product of the deflagration phase is NSE material with about a third of the material
composed of carbon burning products and intermediate mass elements.

These simulations show that through the course of an off-centered ignition, if a
detonation is not triggered as the flame breaks down when it moves through the low density layers
of the star on its way to the surface (i.e. the first scenario in \citep{1997ApJ...478..678K}) and if the
energy release during the deflagration is sufficient to significantly disrupt the
star and cause it to rapidly expand, ash will flow
around the surface (mixing with cold fuel as it flows), and stall in the collision
region while the star is expanding.
As the WD contracts, the conditions in the fuel-ash mixture are pushed to
higher temperature and density until the values exceed that which are necessary
to produce a gradient-triggered detonation.
Once the detonation is triggered in the mixture, it quickly consumes the star.

\section{Comparison of Detonation Mechanism to Previous GCD Models\label{SEC:COMPARE}}

The classical GCD model of \sneia\ is postulated to be possible
if the energy released during the deflagration phase does
not significantly expand the star.
With little stellar expansion, the fuel density in the collision
region is high enough for compression by the in-flowing ash to
increase the temperature in the fuel and ultimately initiate a detonation.
For the purposes of demonstrating this, we restarted the $16b100o8r$
simulation from \gcjiv\ just prior to detonation with the detonation
physics included in simulation.
Figure \ref{fig:gcdcdet} shows the evolution of the $16b100o8r$ classical GCD model
leading up to the detonation.
The top left panel of the figure is at t=1.8s.
It shows the ash as it approached the collision region.
Fuel pushed ahead of the ash increased
in temperature due to compressional heating.
This hot region formed a jet that flowed
both towards and away from the surface of the
star.
The top right panel is at t=1.9s.
The ash converged further into the collision region.
The hot fuel continued to increase in temperature and had
begun to slowly burn carbon as a result.
This smoldering also increased the fuel temperature.
The bottom left panel is at t=2.22 s.
At this time the ash flows have collided.
The hot fuel has further increased in
temperature due to further compressional
heating and due to combustion.
The head of the hot jet has pushed the high temperature
region and has reached the higher density layers of the white
dwarf.
The sufficient thermodynamic conditions to produce a detonation have been
met and a detonation was triggered in the hot smoldering material.
The formation of the detonation can be seen at the head of the hot jet.
Finally, the bottom right panel at t=2.3s shows the simulation
after the detonation occurred.
The detonation wave is propagating outward from its initiation point.
The high temperature region behind the detonation as well as
the smooth detonation front is visible in this image.

The PGCD model of \sneia\ is possible
if the deflagration phase releases enough energy to cause the
WD to expand significantly, but not
so much energy that the star becomes unbound.
Figure \ref{fig:gcdpdet} shows the evolution of the
PGCD model, \gcdsim{168},  leading up to the detonation.
The top left panel is at t=3.5s where the mixture
was beginning to approach the collision region.
In contrast to the classical GCD, the ash flows
and collision region were at low densities and no significant
compressional heating of the fuel in the
collision region occurred.
The top right panel is at t=3.8s.
Note that this panel is zoomed with respect to the previous panel.
The fuel-ash mixture continued to push into the collision region.
The mixture began to heat since the WD was contracting
and the ash was pushing to higher densities
towards the core of the WD.
The bottom left panel is at t=4.1s.
Note that this panel is again zoomed in from the previous panel.
The WD has continued to contract and the mixture
continued to move to higher densities.
The mixture heated significantly due to 1) compressional heating
from the mixture pushing to higher densities and from the
WD contracting and 2) the mixture reaching
conditions that caused it to smolder.
The mixture then approached the critical density above which a detonation
would be initiated.
The bottom left panel is at 4.22s
Note that, again,  this panel is zoomed in from the previous panel.
The mixture reached the necessary thermodynamic conditions for a gradient-triggered
detonation.
The detonation was triggered in the simulation as described in section \ref{subsec:numreal}.
Two detonation waves can be seen emerging from the mixture layer in the image.
The bottom right panel is at t=4.3s.
The detonation is propagating outward from its initiation point through the star.
The high temperature region behind the detonation wave as well as
the smooth detonation front are clearly visible.

In summary, the classical GCD model detonates as the WD is expanding.
The thermodynamic conditions for detonation are reached when the ash flows compress and heat fuel in
the collision region and force this hot fuel into the high density layers of the star.
In contrast, the PGCD model detonates as the WD contracts.
The thermodynamic conditions for detonation are achieved by the combination of the
contraction of the WD which increases the temperature and density of the fuel-ash mixture, 
as the mixture pushes its way towards the high density core of the WD.

\section{Discussion\label{SEC:DISCUSSION}}

\subsection{Properties of the PGCD Scenario}

\subsubsection{Comments on the Detonation Mechanism}
In the classical GCD model, flowing ash must compress fuel in
the collision region to achieve the thermodynamic conditions
necessary to launch a detonation.
This would seem to require the ash flows to converge symmetrically
so that they focus on a localized region.
Thus far, the only successful classical GCD simulations have been
performed in 2D cylindrical geometry (\citep{2009ApJ...693.1188M} for example)
which imposes azimuthal symmetry, and in 3D with single bubble ignition points
as initial conditions which, by their nature, evolve in an azimuthally symmetric
manner.
Our PGCD models were initiated with multiple ignition points.
These ignition points evolved into an asymmetric volume of ash which
produced asymmetric ash flows on the surface.
However, the PGCD depends less stringently on the details of the
ash flows.
As long as the mixture is redirected towards the core of the star,
the contraction of the WD will do the rest, even if
the flow lacks symmetry.
As a result, the PGCD model, in principle, is a more robust
detonation mechanism compare to the classical GCD scenario in that it only depends on
getting the fuel-ash mixture to
a high enough density as the star contracts.

\subsubsection{Nucleosynthesis}
For any \snia\ model, the luminosity of the
model is strongly related to the amount of \iso{Ni}{56}
produced in the explosion.
This depends in turn on the density of the plasma before it is burned.
A fair rule of thumb is that if $\rho \gtrsim 1.0\times 10^{7}$ g/cm$^3$
then that material will burn to NSE, which is predominantly iron-group elements comprised
mostly of \iso{Ni}{56}
(see e.g., the comparison made in figure 12 of \citet{2009ApJ...693.1188M}
and the discussion in their Section 5).
The WD in the PGCD model is contracting
from an expanded state when it detonates.
As a result, there is a reduced amount of high density material.
It is therefore in principle possible for the model to produce range of abundances of \iso{Ni}{56}
- and thus of luminosities - depending on the expanded state of the WD when it
detonates.

The classical GCD models described in \gcjiv\ detonate as the WD
is expanding after a weak deflagration phase.
The star is still fairly compact and there is ample high density material 
when the star detonates.
The amount of \iso{Ni}{56} produced in each of those models is greater that 1 \msolar\ 
which corresponds to overly luminous \sneia.
By contrast, the PGCD model is more easily able to access regions in model space that
correspond to the 0.7 \msolar\ values of \iso{Ni}{56} production --- typical of normal \sneia.
Furthermore, the production of intermediate mass elements between the two scenarios is comparable.
The classical GCD models in \gcjiv\ produced between 0.1 \msolar\ to 0.36 \msolar\ of intermediate
mass elements whereas the PGCD produces between 0.22 \msolar\ and 0.33 \msolar\ of intermediate
mass elements when adding the contributions of \iso{Mg}{24}\ and Si-group elements.

\subsubsection{Observational Features of the Pulsational GCD Model}
The PGCD model has a post detonation geometry similar to
the classical GCD model.
The explosion is the result of a single off-centered detonation.
The location of the detonation is at the antipodal point from where
the star ejected the ash from its interior.
This confers approximate azimuthal symmetry upon the system.
The degree of asymmetry is determined in part by the radial offset of
the detonation trigger.
The larger the offset, the higher degree of asymmetry.
We list the radial offset of the detonation location in our simulations
in table \ref{tab:sims}.

In the PGCD model, the star detonates after the cessation of the deflagration phase.
There is no active flame in the interior of the star at the time of detonation and thus no
compositional inhomogeneities from deflagration ash.
When the detonation occurs, it produces a smooth interior of NSE material
surrounded by a layer of intermediate mass elements.
The intermediate mass elements are surrounded by a shroud of ash produced in 
the deflagration, which is a mixture of intermediate mass elements and NSE material.
This would suggest that elements, such as Fe, would be present in the high-velocity, outer layers
of the ejects and appear as such in the spectra.
The flow of this ash over the surface of the WD produces a clumpy, choppy outer
boundary to the intermediate mass element layer of the remnant.
This is similar to structure suggested by the spectropolarimetry
measurements of \citet{2006ApJ...653..490W, 2007Sci...315..212W}.
Furthermore, very early spectra taken at $\sim 1$ day after the explosion
from the \snia\ 2011fe show that there is O, Mg, Si, S, Ca, and
Fe - which are products of the deflagration in our models - in the outer
most layers of the remnant \citep{2011Natur.480..344N}.
Finally, table \ref{tab:sims} lists the post explosion kinetic energy
for the three simulations.
These energies are $\sim 1 \times 10^{51}$ ergs and are consistent
with observations of \sneia.

\subsection{Comparison to Other Work}
It is interesting to compare our results to the results from the 3D simulations in
\citet{2007ApJ...660.1344R}.
They initialize their simulations with a single spherical bubble, a region of
small bubbles emulating a single bubble with surface perturbations,
and a configuration of bubbles confined to a tear-drop-shaped envelope.
Of the six 3D simulations they performed, two of the WDs
in the simulation became unbound due to the energy released during the deflagration phase.
Of the four simulations in which the WD remained bound, the deflagration energy release
was in the range of $1 \times 10^{50}$ -- $3.3 \times 10^{50}$~ergs, or roughly $\sim$ 20\% -- 60\%
of the binding energy of the WD.
Furthermore, one of their single-bubble ignition model released $2.79 \times 10^{50}$~ergs --- comparable
to our \gcdsim{88}\ simulation (their other single bubble ignition model disrupted the star).
These values of the energy released during the deflagration phase are similar to those presented in our work.
They found that the conditions for detonation are not reached in any of their simulations; however,
they stopped their simulations when the conditions for detonation are not met in the scenario
presented in \gcjiv.
The further evolution of their simulations would have been interesting in light of the results presented here.

\section{Conclusion\label{SEC:CONCLUSION}}

We have conducted a series of numerical experiments simulating the evolution of
a WD star in which we initiate a deflagration at off-center ignition points.
The amount of energy released during the deflagration phase is enough to cause the star to
undergo an energetic pulsation.
As in the classical GCD model, the off-center ignition leads to a plume of material
that is ejected from the star, flows laterally over the stellar surface, and
converges on a collision region at the antipodal point from the ash breakout point.
As the WD contracts, it creates thermodynamic conditions 
in the collision region such that it is possible for the Zel'dovich gradient mechanism
to trigger a detonation.
We find that these necessary (but not sufficient) conditions for detonation are reached in all three of our models.
The energy deposition from the deflagration phase in these models ranges from 38\%  to 78\% of the
binding energy of the WD.
We further comment that the detonation mechanism in the PGCD depends only
on the bulk fluid motion of the system after the deflagration is ignited as opposed to
a specific flow pattern, such as the focusing of the ash flows in the collision region
in the classical GCD.
This property confers a measure of robustness to the detonation mechanism.

Finally, the features of the PGCD qualitatively
agree with the observations of \sneia, insofar as such
comparisons can be made without subjecting the remnant to a radiation
transfer treatment in order to compute actual light curves. 
The three models produced upper limits on the yields of \iso{Ni}{56} ranging from 0.72 \msolar\ to
0.98 \msolar.
These \iso{Ni}{56} yields are less than those produced in the classical GCD
models of \gcjiv, primarily because the WD in our
models is in a more expanded state at the time of detonation and
contains a lower abundance of high-density, NSE-producing matter.
This class of models allows \sneia\ to evolve and detonate
with characteristics similar to the classical GCD while producing
supernovae of lower luminosity.

\begin{figure}
\begin{center}
\includegraphics[width=1.0 \textwidth]{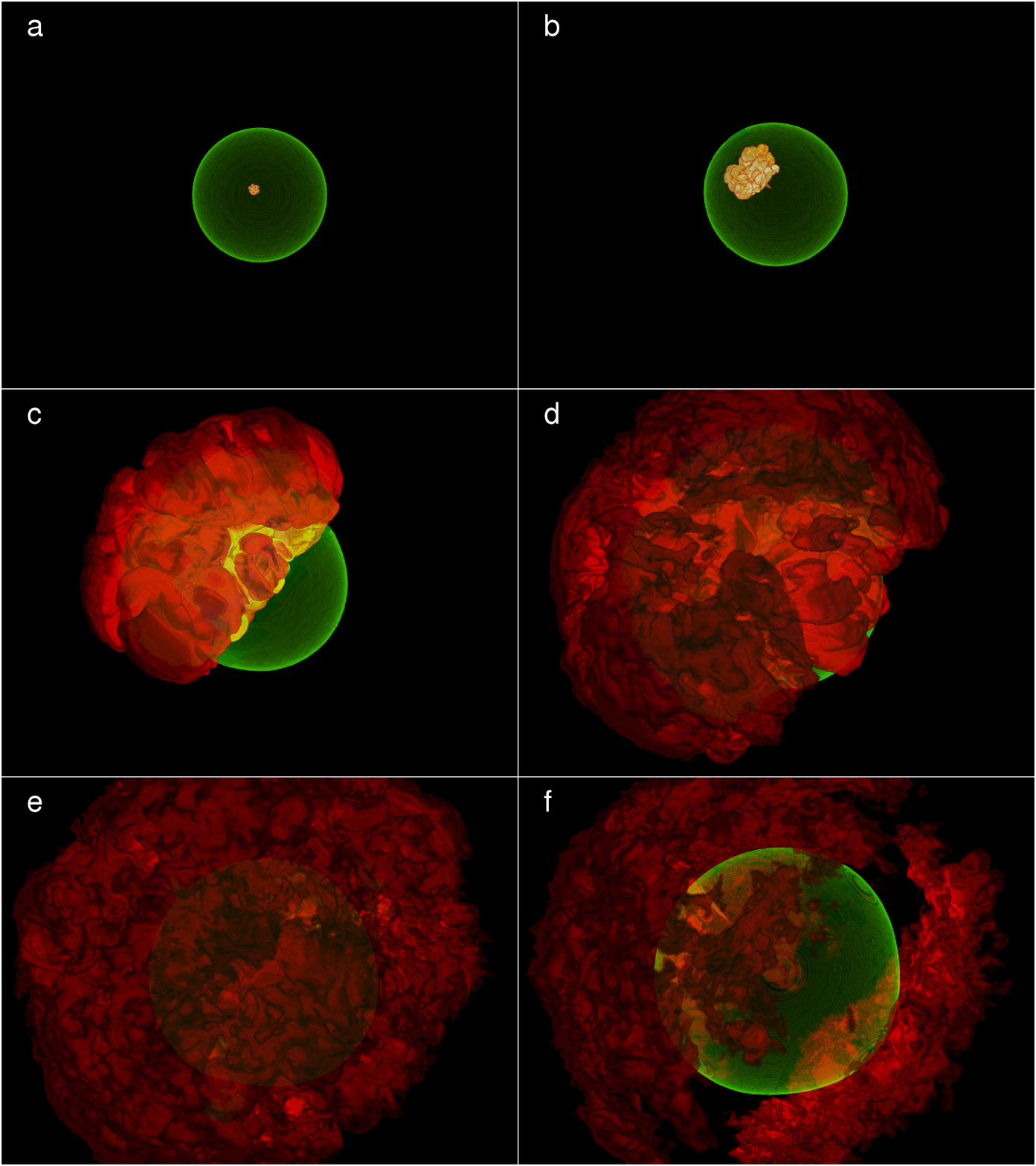}
\caption{
Six snapshots of the \gcdsim{168} simulation.
The green contour approximates the location of the
WD surface.
The range of orange colors are regions
of high temperature.
Bright orange is at the top end of the scale at $5\times 10^9$ K and
dark orange is at the bottom at $1\times 10^9$ K.
The high temperature regions consist primarily hot ash.
The simulation time associated with each panel is:
($a$) 0.0 s, ($b$) 0.6 s, ($c$) 1.1 s, ($d$) 1.4 s,
($e$) 1.9 s, and ($f$) 2.2 s.
Further discussion of this figure can be found in section \ref{subsec:simdescript}.
\label{fig:gcdpBrads1}}
\end{center}
\end{figure}

\begin{figure}
\begin{center}
\includegraphics[width=1.0 \textwidth]{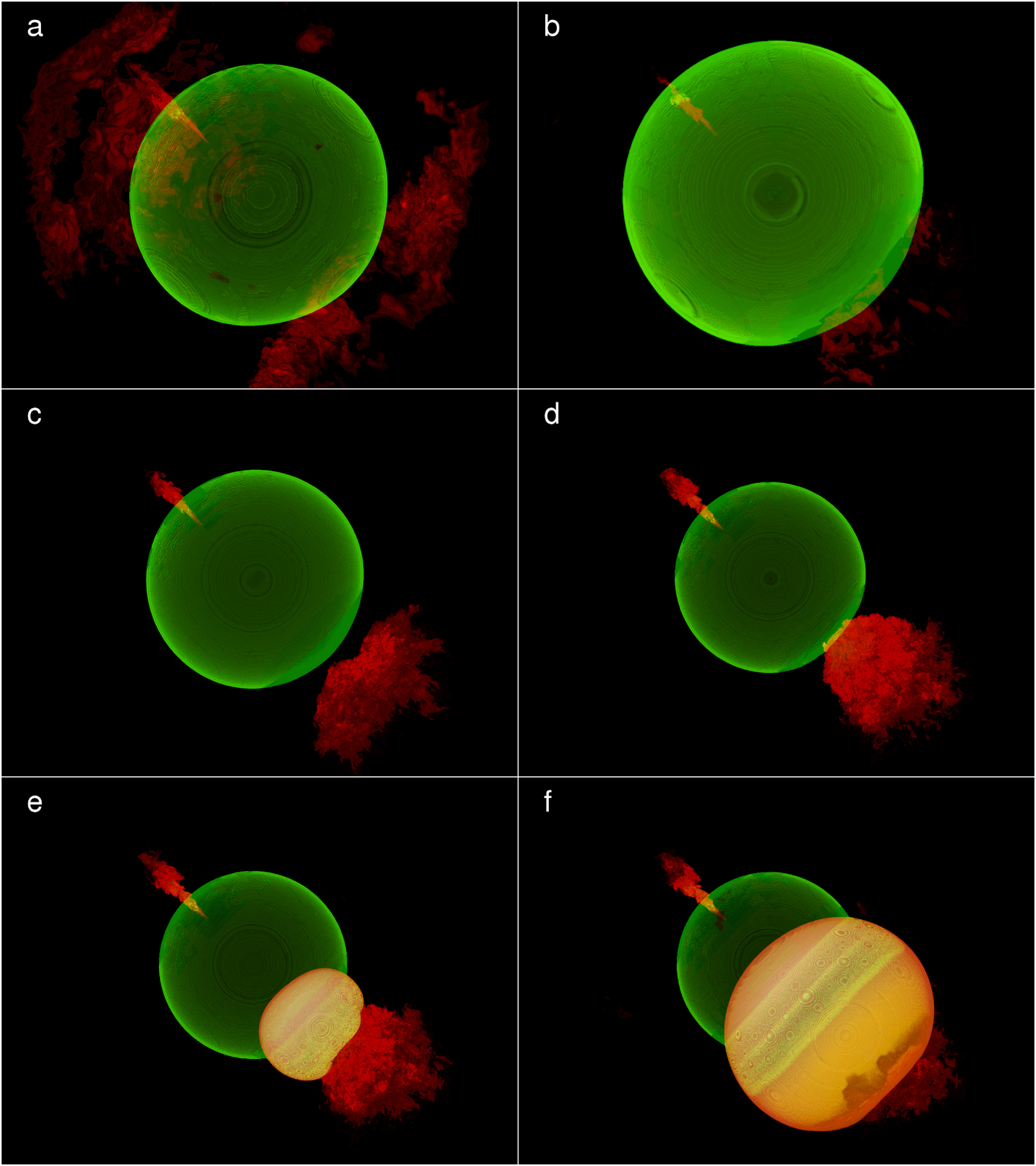}
\caption{
Six snapshots of the \gcdsim{168} simulation.
The green contour approximates the location of the
WD surface.
The range of orange colors are regions
of high temperature.
Bright orange is at the top end of the scale at $5\times 10^9$ K and
dark orange is at the bottom at $1\times 10^9$ K.
The high temperature regions consist primarily hot ash.
The simulation time associated with each panel is:
($a$) 2.5 s, ($b$) 3.1 s, ($c$) 3.7 s, ($d$) 4.2 s,
($e$) 4.3 s, and ($f$) 4.4 s.
Further discussion of this figure can be found in section \ref{subsec:simdescript}.
\label{fig:gcdpBrads2}}
\end{center}
\end{figure}

\begin{figure}
\begin{center}
\includegraphics[width=1.0 \textwidth]{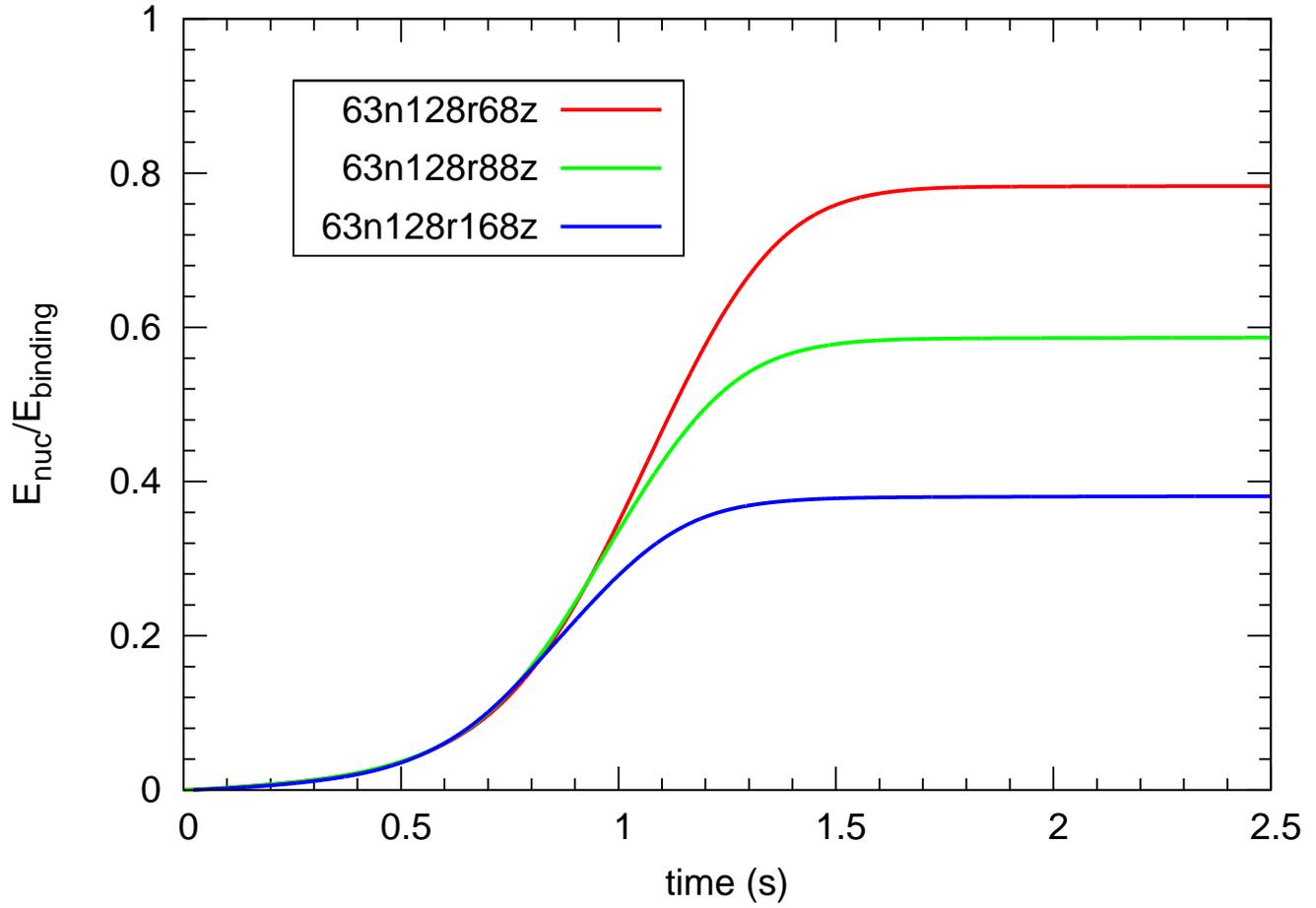}
\caption{The fraction of binding energy increased due to nuclear energy input from the deflagration
phase vs time for the three simulations.
\label{fig:enucVsTime}}
\end{center}
\end{figure}

\begin{figure}
\begin{center}
\includegraphics[width=1.0 \textwidth]{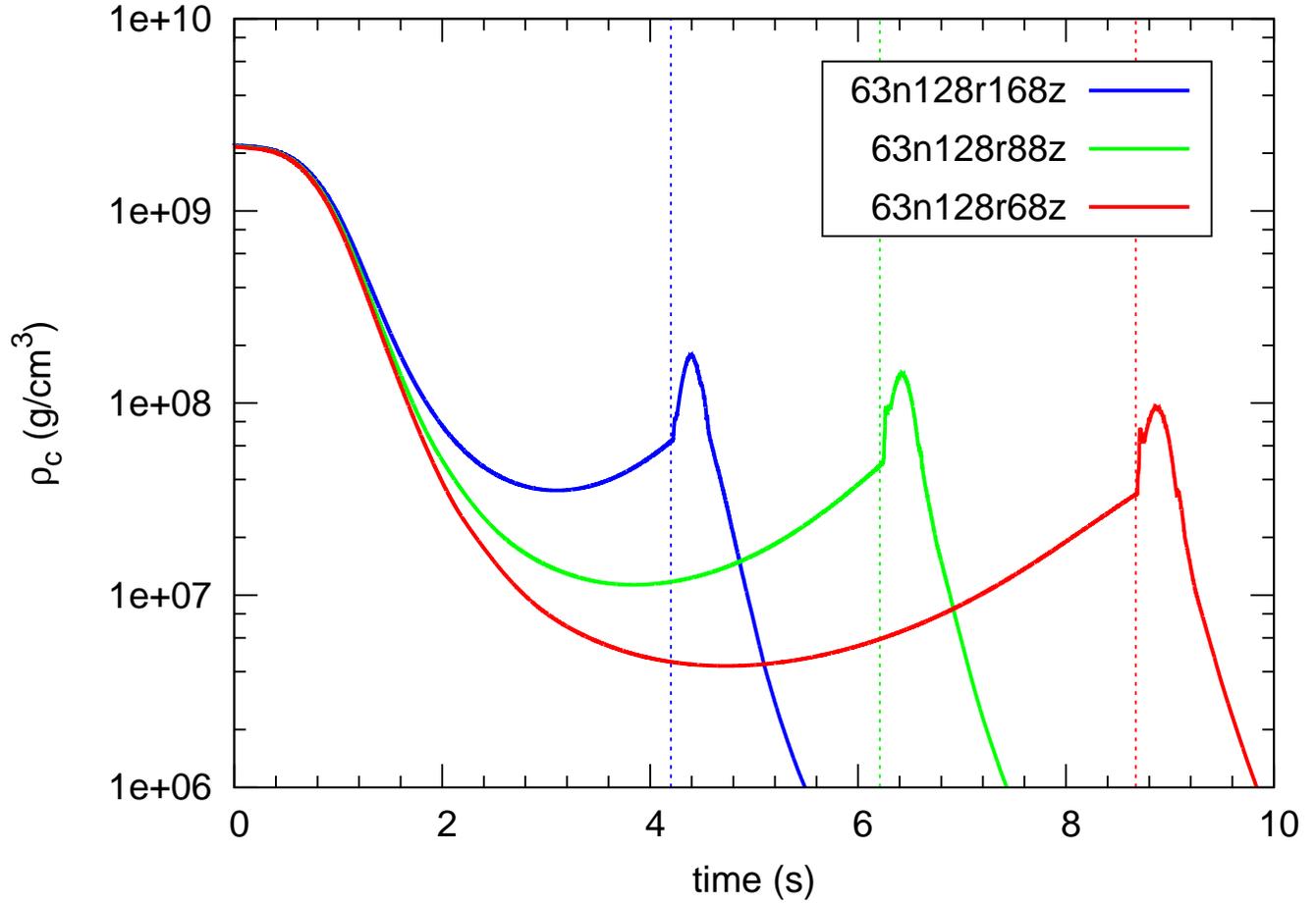}
\caption{The maximum density in the computational domain (which corresponds to the central density, $\rho_{c}$,
of the WD during the pre-detonation phase) vs time for the three simulations.
This plot demonstrates the expansion and contraction of the WD prior to detonation.
The vertical lines mark the time at which a detonation occurred in each simulation.
The sharp rise in density after the detonation is due to density enhancements by
the detonation wave.
\label{fig:rhocVsTime}}
\end{center}
\end{figure}

\begin{figure}
\begin{center}
\includegraphics[width=1.0 \textwidth]{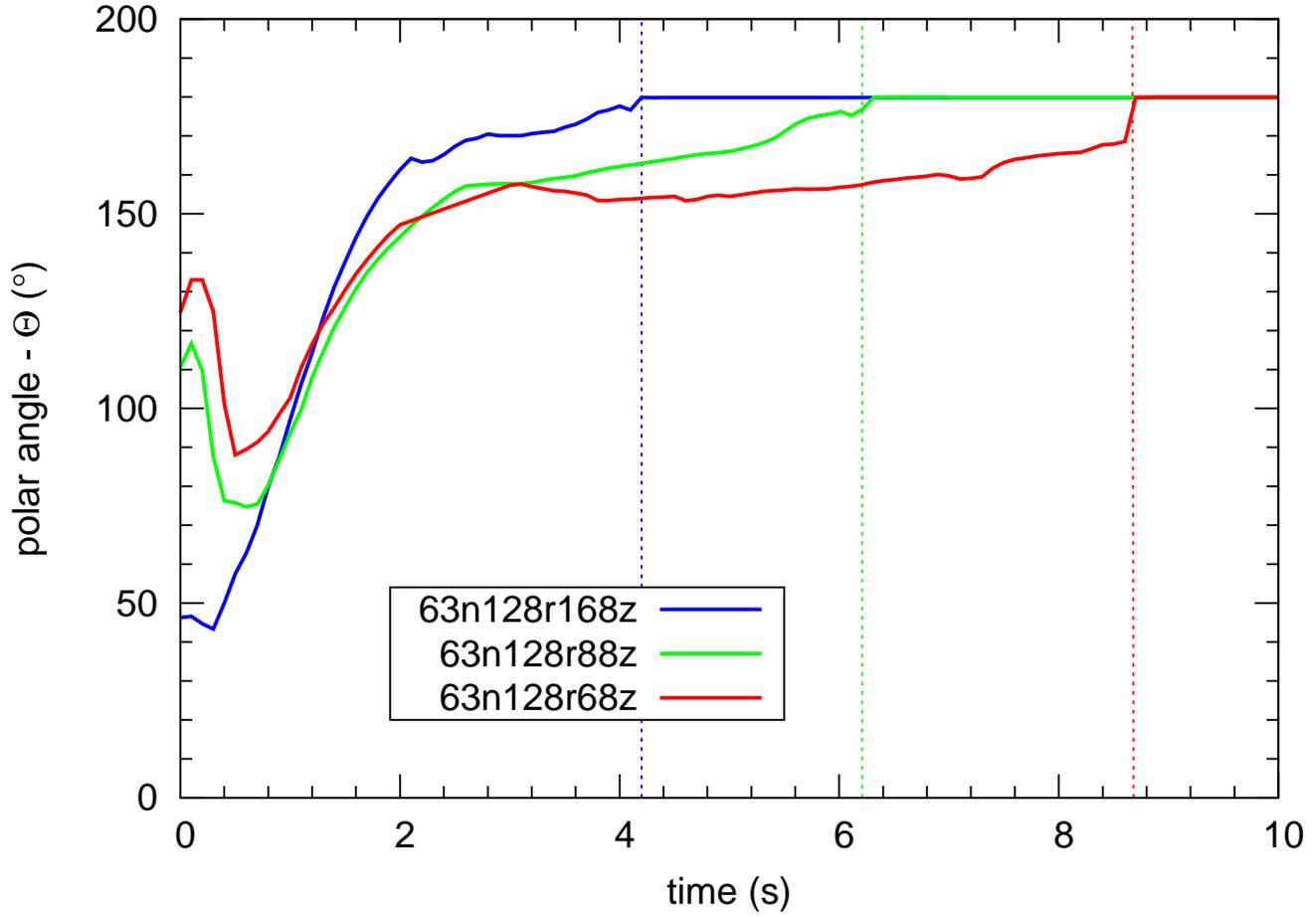}
\caption{The polar angle, $\theta$, of the leading edge of the ash
vs time.
This plot shows the progress of the fuel-ash mixture as it
flows around the WD.
Note that the initially large polar angles during the
first second of evolution is due to the flame burning
into the lower hemisphere of the WD.
The vertical lines mark the time at which a detonation occurred in each simulation. 
\label{fig:thetaAshVsTime}}
\end{center}
\end{figure}

\begin{figure}
\begin{center}
\includegraphics[height=1.0 \textwidth]{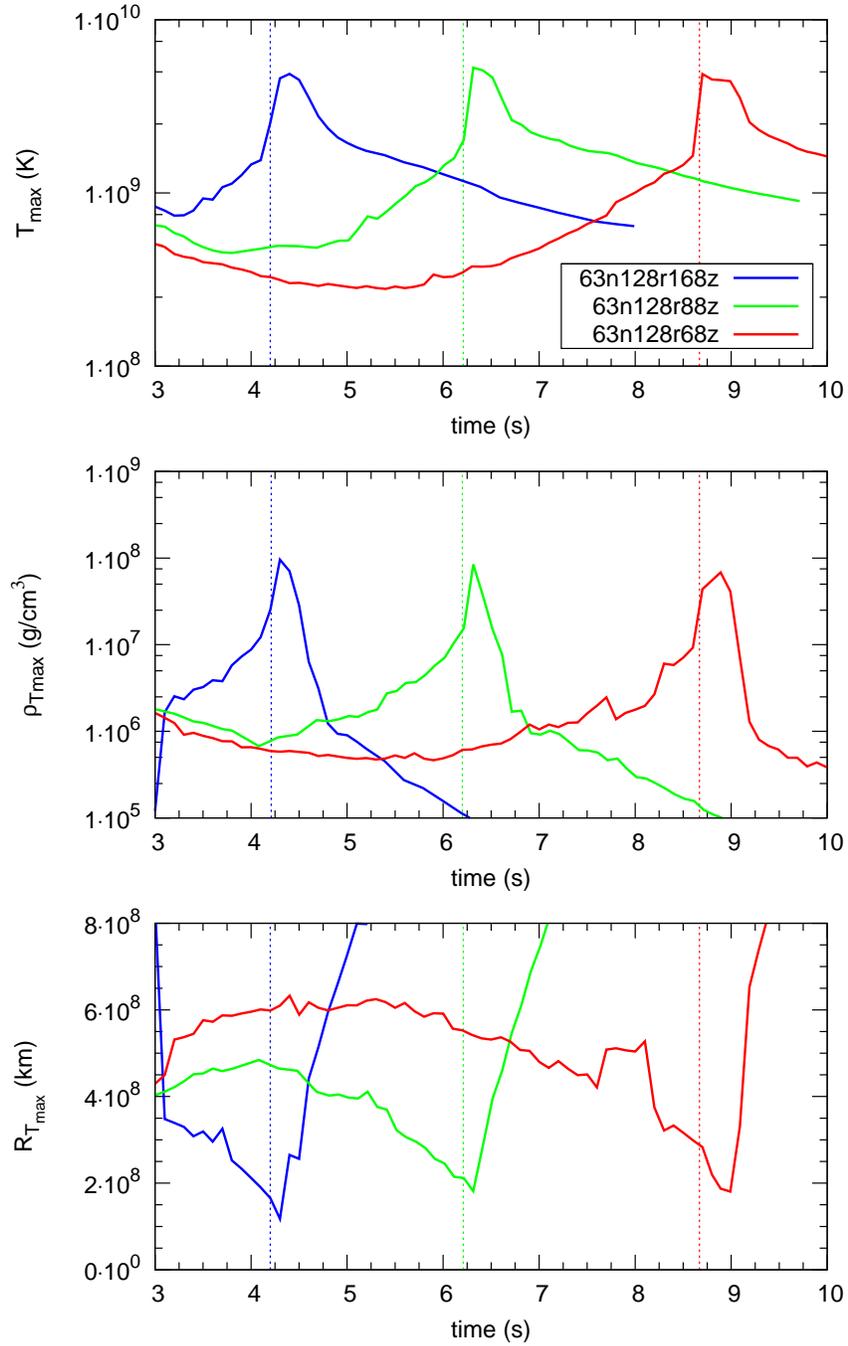}
\caption{Conditions in the collision region:
({\it top}) The temperature of the computational cell with
the maximum temperature in the collision region and
whose material was composed of the fuel-ash mixture.
The maximum temperature is plotted as a function of time.
({\it middle}) The density of the computational cell
selected with the criteria described in the top figure
as a function of time.
({\it bottom}) The radius of the computational cell
selected with the criteria described in the top figure
as a function of time.
The vertical lines mark the time at which a detonation occurred in each simulation.
\label{fig:maxTempVsTime}}
\end{center}
\end{figure}

\begin{figure}
\begin{center}
\includegraphics[width=0.75 \textwidth]{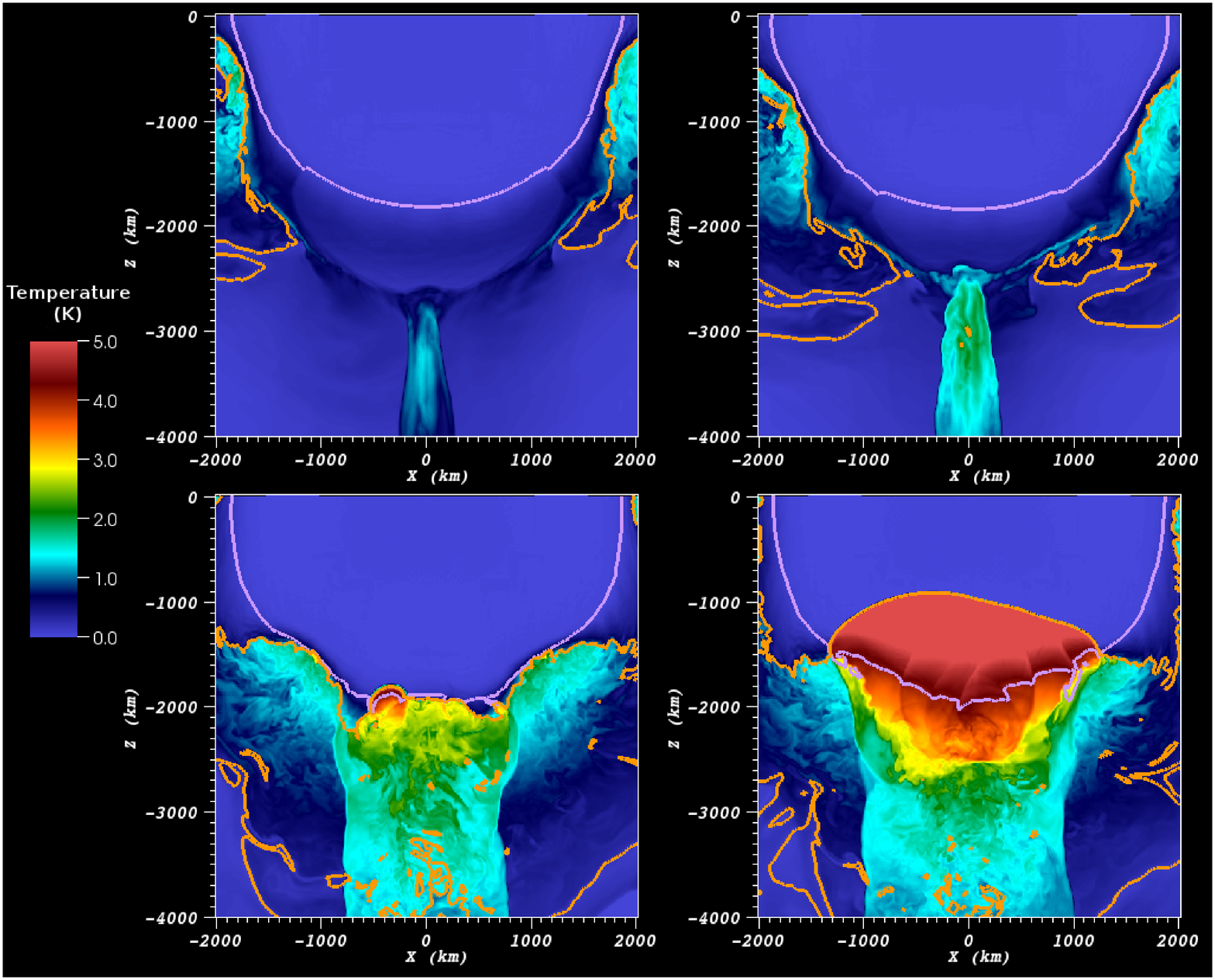}
\caption{
Slice images of the lead up to detonation
of the classical GCD model of \sneia.
This model is the $16b100o8r$ detailed in \gcjiv.
The four images are of the bottom hemisphere of the white
dwarf.
The slice plane is the x-z plane.
The magenta contour is a density contour at $\rho = 1.0 \cdot 10^7$
g/cm$^3$ - the nominal density at which hot C/O would detonate.
The orange contour delineates material that is pure fuel from material
that contains ash (e.g. ash that was converging on the collision region).
The colors indicate temperature whose values correspond to the color bar
on the left.
The color bar is in units of $10^{9}$ K.
({\it top left})
t=1.8s.
Ash approaches the collision region and
a hot region forms.
({\it top right})
t=1.9s.
The ash converges further into the collision region.
The hot region increases in temperature and begins to
smolder.
({\it bottom left})
t=2.22s.
The smoldering fuel has pushed into the region with
$\rho > 1.0 \times 10^{7}$ g/cm$^3$ and a detoantion has just formed.
({\it bottom right})
t=2.3s.
The detonation has spread from where it initially started.
The high temperature region as well as the smooth detonation
front is visible.
\label{fig:gcdcdet}}
\end{center}
\end{figure}

\begin{figure}
\begin{center}
\includegraphics[width=0.75 \textwidth]{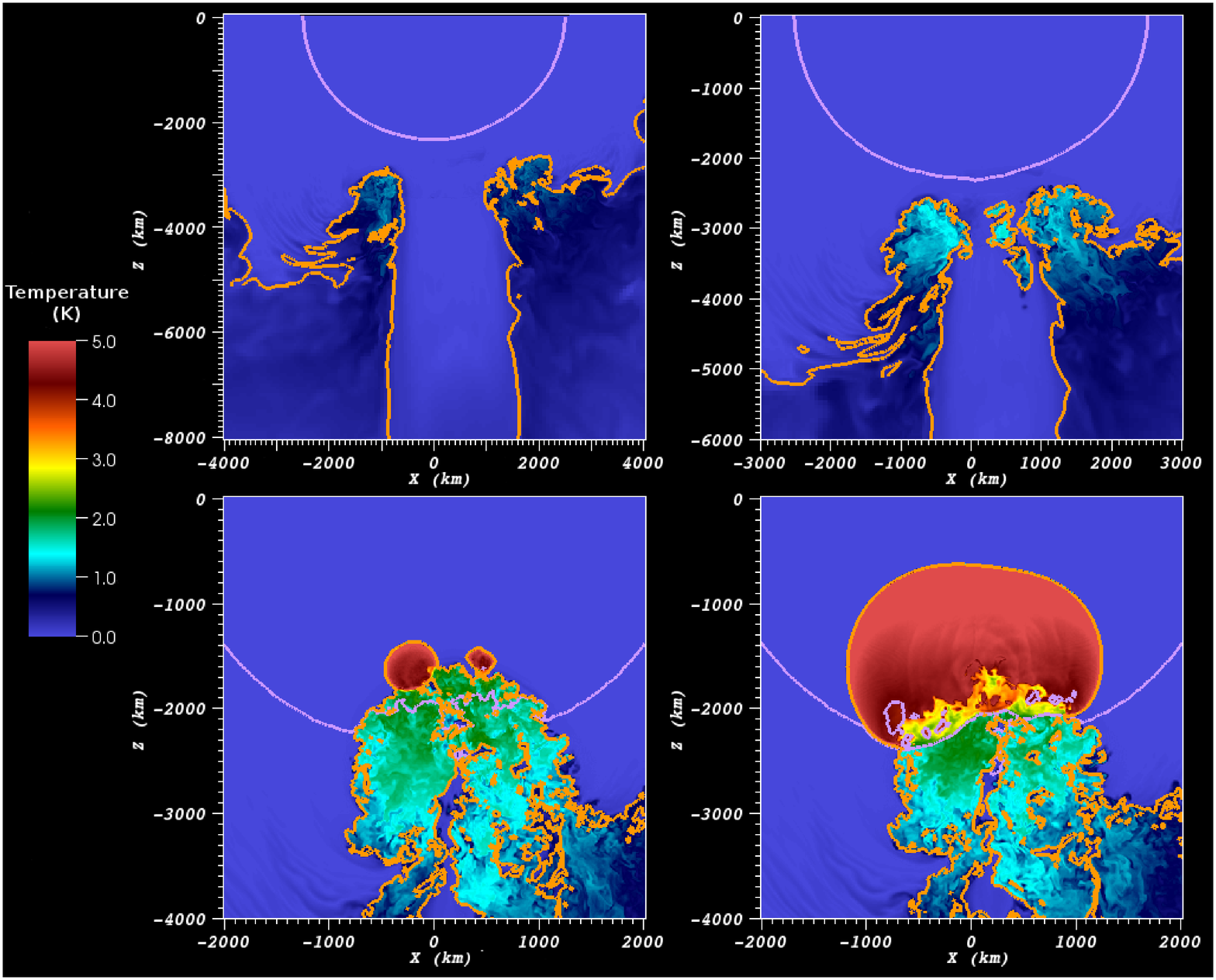}
\caption{
Slice images of the lead up to detonation
of the PGCD model of \sneia.
This model is the \gcdsim{168}.
The four images are of the bottom hemisphere of the white
dwarf.
The slice plane is the x-z plane.
The magenta contour is a density contour at $\rho = 1.0 \cdot 10^7$
g/cm$^3$ - the nominal density at which hot C/O would detonate.
The orange contour delineates material that is pure fuel from material
that contains ash (e.g. ash that was converging on the collision region).
The colors indicate temperature whose values correspond to the color bar
on the left.
The color bar is in units of $10^{9}$ K.
({\it top left})
t=3.5s.
Ash approaches the collision region.
({\it top right})
t=3.8s.
Note that the figure is zoomed in from previous panel.
The fuel-ash mixture continued to push into the collision region.
The mixture heats since the WD was contracting
and the ash was pushing to higher densities
({\it bottom left})
t=4.23s.
Note that the figure is zoomed in from the previous panel.
The mixture has reached the critical temperature and density
and a detonation has just formed at several locations in the
mixture.
({\it bottom right})
t=4.3s.
The detonation wave is spreading through the star.
The high temperature region behind the detonation wave as well as
the smooth detonation front are clearly visible.
\label{fig:gcdpdet}}
\end{center}
\end{figure}

\acknowledgements
The authors thank the FLASH Code Group, especially Chris Daley and Anshu Dubey
for help with the development of and troubleshooting the code.
We thank Brad Gallagher for creating figures \ref{fig:gcdpBrads1} and \ref{fig:gcdpBrads2}.
We also thank Katherine Riley, Mike Papka, and the staff at the Argonne Leadership Computing
Facility at Argonne National Laboratory for help running our large-scale simulations on
Intrepid at ANL.
This work was supported in part at the University of Chicago
by the U.S Department of Energy (DOE) under Contract B523820 to the ASC
Alliances Center for Astrophysical Nuclear Flashes, and in part by the
National Science Foundation under Grant No. AST - 0909132
for the ``Petascale Computing of Thermonuclear Supernova Explosions''.
This research used computational resources awarded under the INCITE program
at ALCF at ANL, which is supported by the Office of Science of the US
Department of Energy under Contract No. DE-AC02-06CH11357.

We dedicate this work to the memory of our colleague and dear friend, Nathan Hearn.

\bibliographystyle{apj}

\clearpage

\clearpage

\clearpage

\clearpage

\clearpage

\clearpage

\clearpage

\clearpage

\clearpage

\begin{deluxetable}{lccccccccccc}
\tabletypesize{\scriptsize}
\tablecaption{List of simulations and their properties\label{tab:sims}}
\tablewidth{0pt}
\tablehead{
\colhead{sim name} &
\colhead{$\Delta x$\tablenotemark{a}} &
\colhead{$n_{ign}$\tablenotemark{b}} &
\colhead{$r_{ign}$\tablenotemark{c}} &
\colhead{$z_{ign}$\tablenotemark{d}} &
\colhead{$E_{nuc,def}$\tablenotemark{e}} &
\colhead{$E_{nuc,def}$\tablenotemark{f}} &
\colhead{$\rho_{c,min}$\tablenotemark{g}} &
\colhead{$t_{\rho_{c,min}}$\tablenotemark{h}} &
\colhead{$t_{det}$\tablenotemark{i}} &
\colhead{$R_{det}$\tablenotemark{j}} &
\colhead{$E_{kinetic}$\tablenotemark{k}} \\
\colhead{} &
\colhead{(km)} &
\colhead{} &
\colhead{(km)} &
\colhead{(km)} &
\colhead{($10^{49}$ ergs)} &
\colhead{($E_{binding}$)} &
\colhead{($10^7$ g/cm$^3$)} &
\colhead{(s)} &
\colhead{(s)} &
\colhead{(km)} &
\colhead{($10^{51}$ $ergs$)}
}
\startdata
\gcdsim{168}  	& 8	&     63	&   128.0	&   168.0	&  18.9		&  0.38		&   3.51	&  3.10		& 4.20        &  1,630        & 1.23		\\
\gcdsim{88}   	& 8	&     63	&   128.0	&   88.0	&  29.1        	&  0.59 	&   1.13	&  3.83		& 6.21        &  2,124        & 1.19		\\
\gcdsim{68} 	& 8	&     63	&   128.0	&   68.0	&  38.6		&  0.78		&   0.43	&  4.71		& 8.67        &  2,660        & 1.05		\\
\enddata
\tablenotetext{a}{Maximum spatial resolution of the simulation.}
\tablenotetext{b}{Number of ignition points.}
\tablenotetext{c}{Radius of the spherical volume containing the ignition points.}
\tablenotetext{d}{Location along the z-axis of the origin of the spherical volume containing the ignition points.}
\tablenotetext{e}{Amount of energy released during the deflagration phase.}
\tablenotetext{f}{Amount of energy released during the deflagration phase as a fraction of the binding energy of the WD.}
\tablenotetext{g}{Central density of the WD at maximum expansion.}
\tablenotetext{h}{Time at which the WD reaches maximum expansion.}
\tablenotetext{i}{Time at which the WD detonates.}
\tablenotetext{j}{Radius of the location where the detonation is initiated in the WD.}
\tablenotetext{k}{Kinetic energy contained in post-explosion nebula.}
\end{deluxetable}

\clearpage

\begin{deluxetable}{lccccc}
\tabletypesize{\scriptsize}
\tablecaption{Final yields from simulations with multiple ignition points.\label{tab:det}}
\tablewidth{0pt}
\tablehead{
\colhead{sim name} &
\colhead{X(\iso{C}{12})} &
\colhead{X(\iso{O}{16})} &
\colhead{X(\iso{Mg}{24})} &
\colhead{X(Si-group)} &
\colhead{X(NSE)}   \\
\colhead{} &
\colhead{\msolar} &
\colhead{\msolar} &
\colhead{\msolar} &
\colhead{\msolar} &
\colhead{\msolar} 
}
\startdata
\gcdsim{168}  & 0.064		& 0.099		& 0.030      	&   0.19        &  0.98        \\
\gcdsim{88}   & 0.091		& 0.14		& 0.045        	&   0.20        &  0.89        \\
\gcdsim{68}   & 0.13		& 0.19		& 0.062        	&   0.27        &  0.72        \\
\enddata
\end{deluxetable}

\begin{deluxetable}{lccc}
\tabletypesize{\scriptsize}
\tablecaption{Deflagration products of simulations with multiple ignition points.\label{tab:def}}
\tablewidth{0pt}
\tablehead{
\colhead{sim name} &
\colhead{X(\iso{Mg}{24})} &
\colhead{X(Si-group)} &
\colhead{X(NSE)}   \\
\colhead{} &
\colhead{\msolar} &
\colhead{\msolar} &
\colhead{\msolar} 
}
\startdata
\gcdsim{168}  & 0.017       &   0.025        &  0.086        \\
\gcdsim{88}   & 0.022       &   0.033        &  0.142        \\
\gcdsim{68}   & 0.030       &   0.044        &  0.190        \\
\enddata
\end{deluxetable}

\end{document}